\documentclass[letter,twocolumn]{jpsj3}

\usepackage{txfonts}
\usepackage{bm}
\usepackage{braket}
\usepackage{color}
\usepackage{cite}
\usepackage{ulem}

\title{ 
Odd-Parity Multipoles by Staggered Magnetic Dipole and Electric Quadrupole Orderings in CeCoSi
}

\author{Megumi Yatsushiro and Satoru Hayami\thanks{Present address: Department of Applied Physics, The University of Tokyo, Tokyo 113-8656, Japan}}
\inst{Department of Physics, Hokkaido University, Sapporo 060-0810, Japan} 

\abst{
We investigate a possibility of odd-parity multipole orderings in a locally noncentrosymmetric tetragonal compound 
CeCoSi. 
By performing symmetrical and microscopic mean-field analyses on a two-orbital tight-binding model, we propose potential odd-parity multipoles hidden in staggered antiferromagnetic and antiferroquadrupole orderings in CeCoSi. 
We show that $3z^2-r^2$ type of the magnetic quadrupole is induced by the staggered magnetic dipole ordering for a large crystal-field splitting between two orbitals, while $xy$ type of the electric toroidal quadrupole is emergent by the staggered electric quadrupole ordering for a small crystal-field splitting.
Furthermore, we discuss a magneto-electric effect and elastic-electric effect
due to the odd-parity multipoles, which will be useful to identify order parameters in CeCoSi.
}

\begin{document}
\maketitle

The breaking of the spatial inversion symmetry has been attracting attention 
in condensed matter physics. 
The effect of the inversion symmetry breaking in crystals is described by an antisymmetric spin-orbit interaction (ASOI) in the form of $\bm{g}(\bm{k})\cdot \bm{\sigma}$ where $\bm{g}(\bm{k})$ is an odd function with respect to the wave vector $\bm{k}$ and $\bm{\sigma}$ is the spin.
The ASOI leads to unconventional physical phenomena, such as a current-induced magnetization which is the so-called Edelstein effect~\cite{edelstein1990spin,furukawa2017observation}, 
noncentrosymmetric superconductivity~\cite{bauer2012non}, and spin Hall effect~\cite{sinova2004universal,Qian2014quantum}.

\begin{figure}[h!]
\centering
\includegraphics[width=88mm]{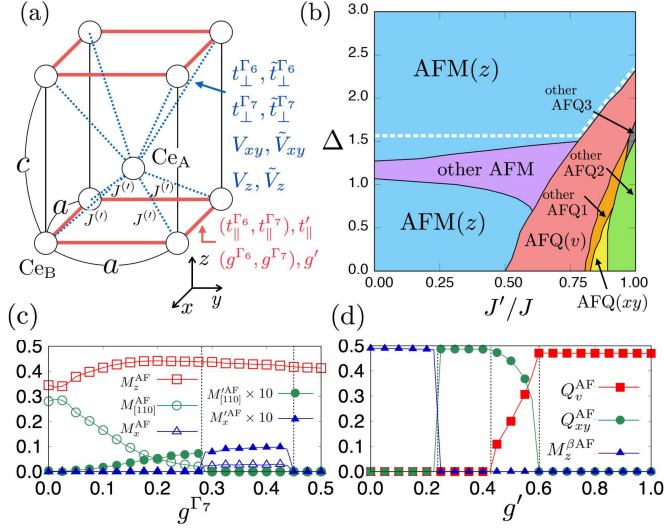}
\caption{
(Color online)
(a) Schematic picture of the crystal structure consisting of Ce$_{\rm A}$ and Ce$_{\rm B}$ with the lattice constants $a$ and $c$.
 (b) The ground-state phase diagram obtained from the mean-field calculations at 
$g^{\Gamma_6}=-0.4$, $g^{\Gamma_7}=0.5$, and $g' = 0.8$.
 AFM($z$) represents the AFM phase with staggered magnetic moments along the $z$ direction.
AFQ($\varv$) and AFQ($xy$) stand for the AFQ phases with $\varv=x^2-y^2$ and $xy$ components of the electric quadrupoles, respectively.
Other AFM and other AFQ1, 2, 3 are the AFM and AFQ phases
characterized by more than one order parameter. 
The phases are metallic (insulating) in the region below (above) the white dashed line.
(c) The intraorbital ASOI dependence of the AFM moments at 
$J'/J=0.2$, $\Delta = 1$, 
$g'=0.8$, and $g^{\Gamma_6}=-0.8g^{\Gamma_7}$.
(d) The interorbital ASOI dependence of the interorbital multipole moments at 
$J'/J=0.7$, $\Delta=0.5$, 
$g^{\Gamma_6}=-0.4$, and $g^{\Gamma_7}=0.5$.
 \label{fig:phase_diagram}
}
\end{figure}

Similar noncentrosymmetric physics arises in a locally noncentrosymmetric system with the staggered-type ASOI once a spontaneous electronic ordering breaks the global inversion symmetry.
For example, staggered magnetic and/or orbital orderings on a zigzag chain~\cite{yanase2013magneto,hayami2015spontaneous,sumita2016superconductivity,hayami2016asymmetric}, honeycomb~\cite{hayami2014toroidal,hayami2014spontaneous,hayami2016emergent,yanagi2018manipulating,takashima_2018nonreciprocal}, diamond~\cite{hayami2019emergent,ishitobi2019magnetoelectric}, and bi-layer structures~\cite{hitomi2014electric, hitomi2016electric,hitomi2019magnetoelectric} give rise to cluster-type odd-parity multipoles, such as the magnetic toroidal dipole and electric octupole~\cite{spaldin2008toroidal,hayami2018classification,watanabe2018group}. 
The $f$-electron metallic compound CeCoSi is a candidate for such cluster-type odd-parity multipoles.
The crystal structure is a centrosymmetric tetragonal CeFeSi-type structure ($P4/nmm$, $D_{\rm 4h}^7$, No.~$129$) and there are two Ce sites (referred as Ce$_{\rm A}$ and Ce$_{\rm B}$) connected by the inversion operation~\cite{bodak1970crystal}, as shown in Fig.~\ref{fig:phase_diagram}(a).  
While changing temperature and pressure, CeCoSi undergoes two phase transitions: 
the antiferromagnetic (AFM) order at $T_{\rm N} =8.8$~K at ambient pressure~\cite{chevalier2004effect,chevalier2006antiferromagnetic} and the hidden order, the latter of which dominantly appears under pressure~\cite{lengyel2013temperature,tanida2018substitution, tanida2019successive}.
Recently, the experiment implies that the hidden order under pressure corresponds to the antiferroquadrupole (AFQ) order~\cite{tanida2018substitution, tanida2019successive}, since it shows similar behavior to the AFQ phase observed in CeB$_6$ and CeTe~\cite{fujita1980anomalous,takigawa1983nmr, effantin1985magnetic,nakamura1994magnetic,sakai1997new,kuramoto2009multipole,kawarasaki2011pressure}.
Interestingly, the unit of the staggered AFM and AFQ orders in CeCoSi accompanies the cluster-type odd-parity multipoles, as the staggered alignment of the even-parity multipoles at two Ce sites breaks the global inversion symmetry.
However, it has not been clarified what types of odd-parity multipoles can be active in the AFM and AFQ phases.
The theoretical identifications are helpful not only to determine order parameters but also to explore physical phenomena driven by odd-parity multipoles~\cite{Khanh2017,saito2018evidence, shiomi2019observation}.

In this Letter, we theoretically investigate cluster-type odd-parity multipoles in CeCoSi.
By examining a two-orbital model including the staggered-type ASOI and crystal-field (CF) splitting on the basis of the group theory and mean-field analyses,
we show that the intraorbital AFM order with $3z^2-r^2$ type of the magnetic quadrupole 
 is stabilized for the large CF splitting, whereas the interorbital AFQ order with $xy$ type of the electric toroidal quadrupole is dominantly realized for the small  CF splitting.
We also study the stability of the AFM and AFQ states in terms of the two types of staggered ASOIs in intraorbital and interorbital spaces.
Furthermore, we discuss the temperature dependence of the magneto-electric effect for each odd-parity multipole ordering.

First, let us describe the local multipole degrees of freedom of the $f^1$ electron configuration in the Ce$^{3+}$ ion~\cite{kusunose2008description,kuramoto2009multipole,santini2009multipolar,suzuki2018first}.
Under the tetragonal CF (the site symmetry $C_{\rm 4v}$), a $J=5/2$ multiplet splits into one $\Gamma_6$ level and two $\Gamma_7$ 
levels.
The active multipoles in the Kramers doublet of $\Gamma_6$ and $\Gamma_7$ levels are magnetic dipoles $(\hat{M}_x, \hat{M}_{y})=(\sigma_x, \sigma_y)$ and $\hat{M}_z=\sigma_z$, where $\sigma_\mu$ 
($\mu=x, y, z$) is the $2\times 2$ Pauli matrix in quasi-spin space. 

Meanwhile, the higher-rank multipoles are active between the $\Gamma_6$ and $\Gamma_7$ levels as interorbital 
degrees of freedom. 
There are eight independent multipoles: four electric quadrupoles 
$\hat{Q}_{\varv}=\tau_x $ ($\varv=x^2-y^2$), $\hat{Q}_{xy}=\sigma_z\tau_y$, and ($\hat{Q}_{yz}, \hat{Q}_{zx})=(\sigma_x\tau_y, \sigma_y\tau_y)$, two magnetic dipoles $(\hat{M}'_x, \hat{M}'_y) = (\sigma_x\tau_x, - \sigma_y\tau_x)$, and two magnetic octupoles $\hat{M}_{xyz} = \tau_y$ and $\hat{M}_z^\beta =  \sigma_z\tau_x$, where $\tau_\nu$ ($\nu=x, y$) is the $2\times 2$ Pauli matrix in $\Gamma_6$-$\Gamma_7$ space~\cite{hayami2018microscopic}. 
The total sixteen multipoles in intraorbital ($\Gamma_6$-$\Gamma_6$ or $\Gamma_7$-$\Gamma_7$) space except for the electric monopole $\sigma_0=1$ and interorbital ($\Gamma_6$-$\Gamma_7$) space, and their irreducible representations at the site symmetry $C_{\rm 4v}$ are summarized in Table~\ref{table:multipole}.

The cluster-type odd-parity multipoles are accompanied with the staggered alignment of the local even-parity multipoles 
at Ce$_{\rm A}$ and Ce$_{\rm B}$ sites~\cite{hitomi2014electric,watanabe2017magnetic,suzuki2019multipole}.
Within the intraorbital $\Gamma_6$ or $\Gamma_7$ space, the staggered orders of $(\hat{M}_x, \hat{M}_y)$ and $\hat{M}_z$ induce magnetic toroidal dipoles $(T_y, T_x)$ and a magnetic quadrupole $M_u$ ($u=3z^2-r^2$), respectively.
On the other hand, the staggered interorbital orders 
of $\hat{Q}_{\varv}$, $\hat{Q}_{xy}$, ($\hat{Q}_{yz}$, $\hat{Q}_{zx}$), ($\hat{M}_x^{{\prime}}$, $\hat{M}_y^{{\prime}}$), $\hat{M}_{xyz}$, and $\hat{M}_z^\beta$ give rise to electric toroidal quadrupoles $G_{xy}$ and $G_{\varv}$, electric dipoles ($Q_y$, $Q_x$), magnetic toroidal dipoles ($T_y$, $T_x$), and magnetic quadrupoles $M_{xy}$ and $M_\varv$, respectively.
The correspondence between the local even-parity multipoles and the cluster odd-parity multipoles is shown in Table~\ref{table:multipole}.

{
\tabcolsep=5pt
\begin{table}[h!]
\centering
\caption{
(Left three columns)
Local even-parity multipoles (MP), irreducible representations (irrep.) at the site symmetry $C_{\rm 4v}$,
 and the matrices  
 $\sigma_\mu \tau_\nu$ ($\mu=x,y,z$, $\nu=x,y$). 
(Right three columns)
Cluster odd-parity MPs (OPMP) by the staggered orderings,
irreps. in the point group $D_{\rm 4h}$, and the magnetic point groups (MPG).
The upper and lower rows represent the active multipoles in the intraorbital $\Gamma_6$-$\Gamma_6$ or $\Gamma_7$-$\Gamma_7$ space and interorbital $\Gamma_6$-$\Gamma_7$ space, respectively.
In the multipole notation, the prefixes E, M, ET, and MT stand for electric, magnetic, electric toroidal, and magnetic toroidal multipoles, and the suffixes D, Q, and O are dipole, quadrupole, and octupole, respectively.
The irrep. is represented by Bethe (Mulliken) description, where the superscript means the spatial inversion (time reversal) property ($+$: even, $-$: odd).
The prefix $m$ in Bethe description represents time-reversal odd.
\label{table:multipole}}
\begin{tabular}{lcclcc}
\hline\hline
\multicolumn{3}{c}{Local multipoles} & \multicolumn{3}{c}{Cluster multipoles}\\
\multicolumn{1}{c}{$C_{\rm 4v}$} & \multicolumn{1}{c}{MP} & $\sigma_\mu \tau_\nu$  & \multicolumn{1}{c}{$D_{\rm 4h}$} & \multicolumn{1}{c}{
OPMP} & MPG 
\\ \hline
$m\Gamma_{2}$(${\rm A}_2^-)$ & $\hat{M}_z$ (MD) & $\sigma_z$ & $m\Gamma_{1}^-$(A$_{\rm 1u}^-$) &  $M_{u}$ (MQ) & $4/m'm'm'$ \\
$m\Gamma_{5}$ (${\rm E}^-)$ & $\hat{M}_x$ (MD) & $\sigma_x$ & $m\Gamma_{5}^-$(E$_{\rm u}^-$) & $T_y$ (MTD) & $mm'm$ \\
& $\hat{M}_y$  (MD) & $\sigma_y$ & & $T_x$ (MTD)& $m'mm$ \\ \hline
$\Gamma_{3}$(${\rm B}_1^+)$ & $\hat{Q}_{\varv}$ (EQ) & $\tau_x$ 
& $\Gamma_{4}^-$(B$_{\rm 2u}^+$) & $G_{xy}$ (ETQ) & $\bar{4}m21'$ \\
$\Gamma_{4}$(${\rm B}_2^+)$ & $\hat{Q}_{xy}$ (EQ) &  $\sigma_z\tau_y$
& $\Gamma_{3}^-$(B$_{\rm 1u}^+$) & $G_{\varv}$ (ETQ) & $\bar{4}2m1'$ \\
$\Gamma_{5}$(${\rm E}^+)$ & $\hat{Q}_{yz}$ (EQ) & $\sigma_x\tau_y$
 & $\Gamma_{5}^-$(E$_{\rm u}^+$) & $Q_y$ (ED) & $mm21'$ \\
& $\hat{Q}_{zx}$ (EQ) & $ \sigma_y\tau_y $
 & & $Q_x$ (ED) & $mm21'$\\ 
$m\Gamma_{3}$(${\rm B}_1^-)$ & $\hat{M}_{xyz}$ (MO) & $\tau_y$
  & $m\Gamma_{4}^-$(B$_{\rm 2u}^-$) & $M_{xy}$ (MQ) & $4'/m'mm'$ \\
 $m\Gamma_{4}$(${\rm B}_2^-)$ & $\hat{M}_{z}^\beta$ (MO) & $ \sigma_z\tau_x$
  & $m\Gamma_{3}^-$(B$_{\rm 1u}^-$) & $M_{\varv}$ (MQ) & $4'/m'm'm$ \\
$m\Gamma_{5}$(${\rm E}^-)$ & $\hat{M}'_x$ (MD) & $\sigma_x\tau_x$
 & $m\Gamma_{5}^-$(E$_{\rm u}^-$) & $T_y$ (MTD) &  $mm'm$\\
 & $\hat{M}'_y$ (MD) & $-\sigma_y\tau_x$
  & & $T_x$ (MTD) & $m'mm$\\ \hline\hline
\end{tabular}
\end{table}
}

Next, we construct the tight-binding model including the above multipole degrees of freedom. 
Following the experimental result that the CF ground state is the Kramers doublet and the first CF excited state is higher by $100$~K~\cite{tanida2019successive}, we adopt the quasi-degenerate two-orbital model where the $\Gamma_7$ level is the ground-state level and the $\Gamma_6$ level is the first excited-state level. 
The Hamiltonian is given by
\begin{align}
\label{eq:model_Hamiltonian}
\hat{\mathcal{H}} &= 
\Delta \sum_{{\bm k} \sigma i} f^\dagger_{{\bm k}i\Gamma_6 \sigma} f_{{\bm k}i \Gamma_6 \sigma} 
+\sum_{\bm k\sigma \mu\nu} \sum_{ij lm} [\varepsilon_{\mu\nu} ({\bm k}) \rho_\mu\tau_\nu]^{lm}_{ij}  
f^\dagger_{{\bm k}i l \sigma} f_{{\bm k}jm \sigma} 
\notag\\
&+\sum_{\bm k\sigma \sigma' \mu \nu} \sum_{ij lm} \left\{\left[{\bm g}
_{\mu \nu} ({\bm k}) + {\bm h}
_{\mu \nu} ({\bm k}) \right] \rho_{\mu} \tau_\nu \right\}^{lm}_{ij}
 \cdot {\bm \sigma}^{\sigma\sigma'}
f^\dagger_{{\bm k}i l \sigma} f_{{\bm k}jm \sigma'}  \notag\\
&+\sum_{
\braket{r,s}}
 \left[ J 
\left( \hat{\bm M}_r^{\Gamma_6} \cdot \hat{\bm M}_s^{\Gamma_6} +  \hat{\bm M}_r^{\Gamma_7} \cdot \hat{\bm M}_s^{\Gamma_7} \right) 
+ J' \hat{\bm X}_r \cdot \hat{\bm X}_s
\right],
\end{align}
where $f^\dagger_{{\bm k}il \sigma}$ ($f_{{\bm k}il \sigma}$) is a creation (annihilation) operator of an electron with the wave vector ${\bm k}$, sublattice $i=$ A, B, orbital $l=\Gamma_6, \Gamma_7$, and quasi-spin $\sigma = \uparrow, \downarrow$.
$\rho_\mu, \tau_\nu$, and $\sigma_\xi$ ($\mu, \nu=0,x,y,z$, $\xi=x,y,z$) are the Pauli matrices in sublattice, orbital, and quasi-spin spaces, respectively. 
The first term in Eq.~(\ref{eq:model_Hamiltonian}) is the CF splitting between the $\Gamma_6$ and $\Gamma_7$ levels. 
The second term is the symmetry-allowed hopping term; 
the intraorbital hoppings, $\varepsilon_{00}({\bm k})$ and $\varepsilon_{0z}({\bm k})$, and the interorbital 
hopping, $\varepsilon_{0x}({\bm k})$, between the same sublattices, and the intraorbital hoppings, 
$\varepsilon_{x0}({\bm k}), \varepsilon_{xz}({\bm k}), \varepsilon_{y0}({\bm k})$, and $\varepsilon_{yz}({\bm k})$, between the different sublattices.
By setting the positions of Ce$_{\rm A}$ and Ce$_{\rm B}$ as $(a/2, a/2, c/2-\theta)$ and $(0,0,0)$ with the lattice constants $a$ and $c$, and using the notations $\varepsilon_{\mu l}({\bm k}) \equiv
[\varepsilon_{\mu 0}({\bm k}) + p(l) \varepsilon_{\mu z}({\bm k})]/2$ where $p(l)=+ 1 (-1)$ for $l=\Gamma_6$ ($\Gamma_7$), each $\varepsilon_{\mu\nu} ({\bm k})$ is given by
$\varepsilon_{0 l}({\bm k}) = t_{\parallel}^{l} (c_{k_xa}+c_{k_ya})$, 
$\varepsilon_{0x}({\bm k}) =  t'_{\parallel} (c_{k_xa} - c_{k_ya})$, 
$\varepsilon_{xl}({\bm k}) =  [t_\perp^l c_{k_zc/2}c_{k_z\theta}+\tilde{t}_{\perp}^l s_{k_zc/2}s_{k_z\theta}]c_{k_xa/2}c_{k_ya/2}$, 
and
$\varepsilon_{yl}({\bm k}) = [t_\perp^l c_{k_zc/2}s_{k_z\theta}-\tilde{t}_{\perp}^l s_{k_zc/2}c_{k_z\theta}]c_{k_xa/2}c_{k_ya/2}$,
where $\cos (\cdots) \equiv c_{\cdots}$ and $\sin (\cdots) \equiv s_{\cdots}$ for simplicity.

The third term in Eq.~(\ref{eq:model_Hamiltonian}) is the spin-dependent hopping term originating from the atomic spin-orbit coupling.
The antisymmetric contribution ${\bm g}
_{\mu \nu} ({\bm k})$ with respect to $\bm{k}$ corresponds to the ASOI, which includes the intraorbital contributions, ${\bm g}_{z0}({\bm k})$ and ${\bm g}_{z z}({\bm k})$, and the interorbital contribution, ${\bm g}_{zx}
({\bm k})$, 
between the same sublattices, which are represented by
\begin{align}
\label{eq:ASOI_1}
{\bm g}_{z l}
({\bm k}) &= g^{l} (-s_{k_y a}, s_{k_x a},0), \\
\label{eq:ASOI_2}
{\bm g}_{zx}
({\bm k}) &= g'(-s_{k_y a}, -s_{k_x a},0), 
\end{align}
where ${\bm g}_{z l}
({\bm k}) \equiv  [{\bm g}_{z0}({\bm k}) +p(l) {\bm g}_{z z}({\bm k})]/2$.
Note that the only staggered component of the ASOI appears due to the presence of the global inversion symmetry. 
The ASOI is microscopically derived from the off-site hybridization with the Co 3$d$ electrons and the atomic spin-orbit coupling. 
Meanwhile, the symmetric spin-dependent hoppings between the different sublattices with the different orbitals, are represented by  
${\bm h}_{xy}
({\bm k}) =\{
{\rm Im}[h_x({\bm k})], 
{\rm Im}[h_y({\bm k})], 
-{\rm Re}[h_z({\bm k})]\}$  
and
${\bm h}_{yy}
({\bm k}) =\{
{\rm Re}[h_x({\bm k})], 
{\rm Re}[h_y({\bm k})], 
{\rm Im}[h_z({\bm k})]\}$, where
$h_x({\bm k}) =(V_{xy} c_{k_zc/2}+ i \tilde{V}_{xy} s_{k_zc/2} ) e^{-i\theta k_z} c_{k_xa/2}s_{k_ya/2}$, 
$h_y({\bm k}) =({V}_{xy} c_{k_zc/2}+ i \tilde{V}_{xy} s_{k_zc/2}) e^{-i\theta k_z} s_{k_xa/2}c_{k_ya/2}$, 
and
$h_z({\bm k}) =({V}_{z} c_{k_zc/2}+ i \tilde{V}_{z} s_{k_zc/2}) e^{-i\theta k_z} s_{k_xa/2}s_{k_ya/2}$.

The fourth term in Eq.~(\ref{eq:model_Hamiltonian}) represents the effective antiferroic interactions between the intraorbital multipoles $J >0$ and interorbital multipoles $J'>0$. 
The summation is taken for the four nearest-neighbor A and B sites $\braket{r,s}$,
 as shown in Fig.~\ref{fig:phase_diagram}(a).
$\hat{\bm M}^{l}_{r}
=\frac{1}{2}\sum_{\sigma\sigma'}{\bm \sigma}^{\sigma\sigma'}f_{rl\sigma}^\dagger f_{rl\sigma'}$ ($l= \Gamma_6, \Gamma_7$) 
and $\hat{X}_{r} =\frac{1}{2}\sum_{lm}\sum_{\sigma\sigma'}(\tau_\mu\sigma_\nu)^{lm}_{\sigma\sigma'} f_{rl\sigma}^\dagger f_{rm\sigma'}$
are the magnetic dipole and the eight interorbital multipoles (Table~\ref{table:multipole}) 
at site $r$, respectively, where $f_{rl\sigma}^\dagger$ ($f_{rl\sigma}$) is the Fourier transform of $f^\dagger_{{\bm k}il \sigma}$ ($f_{{\bm k}il \sigma}$).
We adopt the isotropic exchange interactions $J$ and $J'$, which are introduced to mimic
the strong intraorbital and interorbital Coulomb interaction without the spin-orbit coupling~\cite{kugel1973crystal}. 
The intraorbital interaction $J$ favors the AFM ordering, while the interorbital interaction $J'$ favors the antiferroic interorbital multipole orderings, such as the AFQ ordering. 
We note that the intraorbital states with $(\hat{M}_x, \hat{M}_y, \hat{M}_z)$ and interorbital states with $(\hat{Q}_{\varv}, \hat{Q}_{xy}, \hat{Q}_{yz}, \hat{Q}_{zx}, \hat{M}_{xyz}, \hat{M}_z^{\beta},\hat{M}'_{x}, \hat{M}'_{y})$ are degenerate within the $J$ and $J'$ terms, respectively.
We here focus on the effect of the staggered ASOIs in Eqs.~(\ref{eq:ASOI_1}) and (\ref{eq:ASOI_2}) on the stability of each multipole order.

We investigate the ground-state phase diagram of the model in Eq.~(\ref{eq:model_Hamiltonian}) by mean-field calculations.
We use the Hartree approximation for the two-body terms and consider supercells consisting of $80^3$ copies of the two sublattices under the periodic boundary conditions. 
The numerical error of the self-consistent calculations is less than $10^{-4}$.
We adopt the $f^{1}$ configuration, i.e., the $1/4$ filling, and set parameters
 $t_\parallel^{\Gamma_6}=0.8, t_\parallel^{\Gamma_7}=1
 $, $t'_\parallel=0.1$, $t_\perp^{\Gamma_6}=t_\perp^{\Gamma_7} =0.15$, $\tilde{t}_\perp^{\Gamma_6}=\tilde{t}_\perp^{\Gamma_7} =0.05$, ${V}_{xy}=0.15$, $\tilde{V}_{xy}=0.05$, ${V}_z=0.3$, $\tilde{V}_z = 0.1$,  
 $J=2.5$, and $c/a=1.4$.   
We set $\theta=0$. Although $\theta$ is finite in CeCoSi, the effect of nonzero $\theta$ is taken into account for the hopping and interaction parameters along the $z$ direction. 

Figure~\ref{fig:phase_diagram}(b) shows the ground-state phase diagram by changing
$J'/J$ and $\Delta$ for $g^{\Gamma_6}=-0.4$, $g^{\Gamma_7}=0.5$, and $g' = 0.8$.
For large $\Delta$ where the $\Gamma_6$ level is well-separated from the $\Gamma_7$ level, the 
intraorbital multipole instability occurs and the AFM state is stabilized through the intraorbital interaction $J$. 
In a large portion of the AFM regions, the magnetic moments are along the $z$ direction, where
we denote the phase as AFM($z$). This phase is accompanied with the magnetic quadrupole $M_u$, as shown in Table~\ref{table:multipole}.
In the phase diagram, another AFM phase denoted as other AFM is realized around $0.8 \lesssim \Delta \lesssim 1.4 $, where the staggered magnetic moments are tilted from the $z$ direction.
The obtained AFM phases for $\Delta \lesssim 1.56$ are metallic, 
whereas the AFM($z$) phase for $\Delta \gtrsim 1.56$ is insulating.

The magnetic anisotropy in the AFM phases results from the interplay between two types of ASOIs.
Especially, the AFM($z$) state stabilized in the insulating region for large $\Delta$ is presumably 
owing to the intraorbital ASOI. 
Note that a similar tendency is obtained in magnetic insulators in the strongly correlated regime where the effective out-of-plane anisotropic interaction appears~\cite{hayami2016asymmetric}.
In the metallic region, although the effective interaction by the ASOIs is affected by the band structure and must be more complicated, the mean-field results indicate that
the intraorbital ASOI tends to stabilize the 
AFM($z$) state, whereas the interorbital ASOI, whose effect becomes important for large $J'/J$, tends to stabilize the other AFM state with the in-plane moments, as discussed below.

The AFM states are replaced with the AFQ states by decreasing $\Delta$ and increasing $J'/J$ with a finite jump of order parameters.
This is ascribed to the quasi-orbital degeneracy between the $\Gamma_7$ and $\Gamma_6$ levels, whose instability is also found in excitonic states in the multi-orbital $d$-electron systems~\cite{kunevs2014excitonic,kunevs2014excitonicPr}.
The dominant AFQ instability in Fig.~\ref{fig:phase_diagram}(b) is the $Q_\varv$ channel with the electric toroidal quadrupole $G_{xy}$.
The other AFQ states denoted as AFQ($xy$) and other AFQ1, 2, 3 in $J'/J \gtrsim 0.8$ are characterized by the staggered orders of $Q_{xy}$, and linear combinations of ($Q_{xy}, Q_\varv$), ($Q_{xy}, M_{xyz}$), and ($Q_\varv, Q_{xy}, M_{xyz}, M_z^\beta$), respectively.
The stability of these interorbital ordered states is affected by the interplay between two types of ASOIs and the interorbital hopping, as discussed below. 
All the AFQ phases are metallic.

To examine the effect of the ASOI on the AFM($z$) 
state obtained in Fig.~\ref{fig:phase_diagram}(b), we show the intraorbital staggered ASOI $g^{\Gamma_7}$ dependence of the staggered AFM moments while keeping $g^{\Gamma_6} = -0.8 g^{\Gamma_7}$ at $J'/J=0.2$, $\Delta = 1$, and $g'=0.8$ in Fig.~\ref{fig:phase_diagram}(c).
We compute the $\mu$ component of the AFM moment $M_\mu^{\rm AF} \equiv [(M_\mu^{\Gamma_6 {\rm AF}})^2 + (M_\mu^{\Gamma_7 {\rm AF}})^2]^{1/2}$ for $\mu=x,y,z$ and $M_{[110]}^{\rm AF} =
[(M_x^{{\rm AF}})^2 + (M_y^{{\rm AF}})^2]^{1/2}
$ where the staggered component of multipoles $X$ is defined as $X^{\rm AF} = ({X}_{\rm A} -{X}_{\rm B})/2$.
Note that there is also interorbital contribution $M_{x(y)}^{\prime{\rm AF}}$ for the in-plane moments.

In Fig.~\ref{fig:phase_diagram}(c), the AFM($z$) phase is stabilized at $g^{\Gamma_7}=0.5$, as shown in Fig.~\ref{fig:phase_diagram}(b). 
While decreasing $g^{\Gamma_7}$, ${\bm M}^{\rm AF}$ is tilted from the $z$ axis toward the $[100]$ direction 
for $g^{\Gamma_7}\lesssim 0.45$, although $M_{z}^{\rm AF}$ is larger than $M_{x(y)}^{\rm AF}$ and $M_{x(y)}^{\prime{\rm AF}}$. 
The appearance of $M_{x(y)}^{\rm AF}$ and $M_{x(y)}^{\prime{\rm AF}}$ corresponds to the emergence of
the magnetic toroidal dipole $T_y$($T_x$).  
With a further decrease of $g^{\Gamma_7}$, the in-plane moment direction
changes from the $[100]$ to $[110]$ direction at $g^{\Gamma_7}\sim0.275$.
Then, $M^{\rm AF}_{[110]}$ increases while decreasing $g^{\Gamma_7}$ and becomes comparable to $M_z^{\rm AF}$ at $g^{\Gamma_7}=0$, whereas $M^{\prime{\rm AF}}_{[110]}$ is suppressed when decreasing $g^{\Gamma_7}$. 
The result indicates that the intraorbital ASOI favors the AFM($z$) state. 
On the other hand, it also indicates that the AFM state with the in-plane magnetic moments, such as the other AFM state, can be stabilized by the interorbital ASOI~\cite{hayami2015spontaneous,yatsushiro2019atomic}.

Next, we show the effect of the interorbital  
ASOI $g'$ on the AFQ($\varv$) state at $J'/J=0.7$, $\Delta = 0.5$, $g^{\Gamma_7}=0.5$, and $g^{\Gamma_6} = -0.4$.
Figure~\ref{fig:phase_diagram}(d) shows that four interorbital states are stabilized while changing $g'$.
The AFQ($\varv$) phase is stabilized for $0.6 \lesssim g' \lesssim 1$, the other AFQ1 phase is stabilized for $0.425 \lesssim g' \lesssim 0.6$, the AFQ($xy$) phase is stabilized for $0.225 \lesssim g' \lesssim 0.425$, and the staggered $M_z^\beta$ phase appears for $0 \lesssim g' \lesssim 0.225$. 
From the numerical result, the interorbital ASOI $g'$ tends to favor the AFQ($\varv$) state.
On the other hand, the stability of the AFQ($xy$) and the staggered $M_z^\beta$ states for small $g'$ depends on the intraorbital ASOI $g^{\Gamma_6}$ and $g^{\Gamma_7}$ and the interorbital hopping $t'_{\parallel}$. 
The large $g^{\Gamma_7}$ and $g^{\Gamma_6}$ tend to favor the AFQ($xy$) state for small $g'$, while $t'_{\parallel}$ tends to stabilize the $M_z^\beta$ state.
Thus, the stability of the interorbital phases is affected by the competing factors
$g'$, $g^{\Gamma_6}$, $g^{\Gamma_7}$, and $t'_{\parallel}$~\cite{comment_interorbital}.

Finally, we discuss physical phenomena driven by the odd-parity multipoles, which will be helpful to identify order parameters.
We focus on the cross-correlation phenomena 
where the multipole $Y_{\mu}$ is induced by an electric field $E_{\nu}$ as $Y_{\mu}=\sum_{\nu}\chi_{\mu\nu}E_{\nu}$ in the AFM($z$) and AFQ($\varv$) phases, which are dominantly stabilized in the phase diagram in Fig.~\ref{fig:phase_diagram}(b).
The  tensor $\chi_{\mu\nu}$ is calculated by the linear response theory~\cite{yanase2013magneto, watanabe2017magnetic,hayami2018classification} as
\begin{align}
\label{eq:CC}
\chi_{\mu\nu} & = 
 \sum_{\bm k} \sum_{pq} \Pi_{pq}({\bm k}) Y_{\mu{\bm k}}^{pq}v_{\nu{\bm k}}^{qp}= \chi_{\mu \nu}^{\rm (J)} + \chi_{\mu \nu}^{\rm (E)},
\end{align}
where 
$\Pi_{pq}({\bm k}) =e\hbar \{f[\varepsilon_p({\bm k})] - f[\varepsilon_q({\bm k})]\}/\{ Vi[\varepsilon_p({\bm k})-\varepsilon_q({\bm k})][\varepsilon_p({\bm k})-\varepsilon_q({\bm k})+i\hbar \delta]\}$ 
with the eigenenergy $\varepsilon_p({\bm k})$ and the Fermi distribution function $f[\varepsilon_p({\bm k})]$. 
$e$ is the electron charge, $\hbar = h/2\pi$ is the Plank constant, $V$ is the system volume, and $\delta$ is the broadening factor. 
We take $e=\hbar=1$ and $\delta=0.1$.
$Y^{pq}_{\mu {\bm k}}= \braket{p{\bm k}|\hat{Y}_\mu|q{\bm k}}$
and $v_{\nu{\bm k} }^{pq}=\braket{p{\bm k}|\hat{v}_{\nu{\bm k}}|q{\bm k}}$ 
are the matrix elements of the multipole $\hat{Y}_\mu$ and velocity $\hat{v}_{\mu{\bm k}} = \partial \hat{\mathcal{H}}/(\hbar \partial k_\mu)$.
The tensor $\chi_{\mu\nu}$ in Eq.~(\ref{eq:CC}) consists of the dissipative part $\chi_{\mu\nu}^{\rm (J)}$ (current driven part) from the intraband contribution and the non-dissipative part $\chi_{\mu \nu}^{\rm (E)}$ (electric-field driven part) from the interband contribution~\cite{watanabe2017magnetic,hayami2018classification}.

When $\hat{Y}_{\mu}$ is the magnetic dipole $\hat{M}_\mu$, $\chi_{\mu\nu}$ corresponds to the magneto-electric tensor $\alpha_{\mu\nu}$, where the magnetization $M_\mu$ is induced by the electric field $E_{\nu}$
for $\mu, \nu = x,y,z$.
Note that the magneto-electric tensor $\alpha_{\mu\nu}$ consists of three contributions of $\alpha_{\mu\nu}^{(\Gamma_6)}$, $\alpha_{\mu\nu}^{(\Gamma_7)}$, and $\alpha_{\mu\nu}^{\prime}$, as there are three types of
magnetizations $M_\mu^{\Gamma_6}$, $M_\mu^{\Gamma_7}$, and $M_\mu^\prime$ 
in Table~\ref{table:multipole}. 
On the other hand, when $\hat{Y}_{\mu}$ is the electric quadrupole $\hat{Q}_\mu$, $\chi_{\mu\nu}$ is the elastic-electric (inverse piezo-electric) tensor $d_{\mu\nu}$, where the symmetric distortion $\epsilon_{\mu}$ ($\mu=u,\varv,yz,zx,xy$) is induced by $E_{\nu}$. 
As nonzero tensor components correspond to the emergent of odd-parity multipoles, the different types of responses are obtained in each multipole phase, as summarized in Table~\ref{table:cross_correlation}.
In the following, we focus on the behavior of the magneto-electric tensor $\alpha_{\mu\nu}$ in the AFM($z$) and AFQ($\varv$) states.

\begin{table}[h!]
\centering
\caption{
Nonzero components of the magneto-electric ($\alpha_{\mu\nu}$) and elastic-electric ($d_{\mu\nu}$)  
tensors in each 
multipole (MP) phase. 
The magnetic point groups (MPG) and the 
odd-parity multipoles (OPMP)
are also shown.
\label{table:cross_correlation}}
\begin{tabular}{ccccc}\hline \hline
MPG &  MP & 
OPMP & $\alpha_{\mu\nu}$ & $d_{\mu\nu}$ \\\hline
$\bar{4}m21'$ & $\hat{Q}_\varv$ & $G_{xy}$ & $\alpha_{yx}^{\rm (J)}=\alpha_{xy}^{\rm (J)}$ & $d_{zxx}^{\rm (E)} = -d_{yzy}^{\rm (E)}, d_{\varv z}^{\rm (E)}$ \\
$\bar{4}2m1'$ & $\hat{Q}_{xy}$ & $G_{\varv}$ & $\alpha_{xx}^{\rm (J)}=-\alpha_{yy}^{\rm (J)}$ & $d_{yzx}^{\rm (E)}=d_{zxy}^{\rm (E)}$, $d_{xyz}^{\rm (E)}$ \\
$mm21'$ & $\hat{Q}_{yz}$ & $Q_y$ & $\alpha_{zx}^{\rm (J)}$, $\alpha_{xz}^{\rm (J)}$ & $d_{xyx}^{\rm (E)}$, $d_{uy}^{\rm (E)}$, $d_{\varv y}^{\rm (E)}$, $d_{yzz}^{\rm (E)}$\\
$mm21'$ & $\hat{Q}_{zx}$ & $Q_x$ & $\alpha_{zy}^{\rm (J)}$, $\alpha_{yz}^{\rm (J)}$ & $d_{ux}^{\rm (E)}$, $d_{\varv x}^{\rm (E)}$, $d_{xyy}^{\rm (E)}$, $d_{zxz}^{\rm (E)}$ \\ 
$mm'm$ & $\hat{M}_x$ & $T_y$ & $\alpha_{zx}^{\rm (E)}$, $\alpha_{xz}^{\rm (E)}$ & $d_{xyx}^{\rm (J)}$, $d_{uy}^{\rm (J)}$, $d_{\varv y}^{\rm (J)}$, $d_{yzz}^{\rm (J)}$\\
$m'mm$ & $\hat{M}_y$ & $T_x$ & $\alpha_{zy}^{\rm (E)}$, $\alpha_{yz}^{\rm (E)}$ & $d_{ux}^{\rm (J)}$, $d_{\varv x}^{\rm (J)}$, $d_{xyy}^{\rm (J)}$, $d_{zxx}^{\rm (J)}$ \\
$4/m'm'm'$ & $\hat{M}_z$ & $M_u$ & $\alpha_{xx}^{\rm (E)}=\alpha_{yy}^{\rm (E)}$, $\alpha_{zz}^{\rm (E)}$ & $d_{yzx}^{\rm (J)}=-d_{zxy}^{\rm (J)}$ \\
$4'/m'mm'$ & $\hat{M}_{xyz}$ & $M_{xy}$ & $\alpha_{yx}^{\rm (E)}=\alpha_{xy}^{\rm (E)}$ & $d_{zxx}^{\rm (J)}=-d_{yzx}^{\rm (J)}$, $d_{\varv z}^{\rm (J)}$ \\
$4'/m'm'm$ &$\hat{M}_z^\beta$ & $M_\varv$ & $\alpha_{xx}^{\rm (E)} = -\alpha_{yy}^{\rm  (E)}$ & $d_{yzx}^{\rm (J)}=d_{zxy}^{\rm (J)}$, $d_{xyz}^{\rm (J)}$ \\ \hline
\end{tabular}
\end{table}

\begin{figure}[h!]
\centering
\includegraphics[width=88mm]{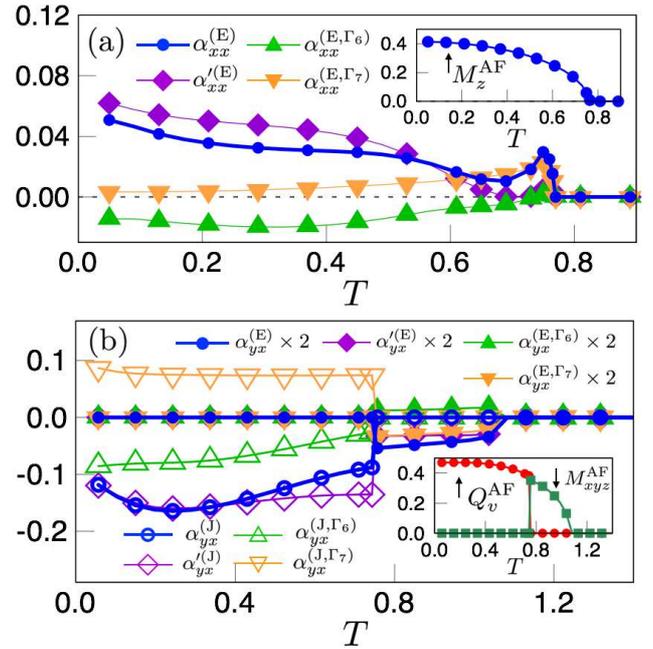}
\caption{(Color online)
(a), (b) Temperature dependences of  
magneto-electric tensors in (a) the AFM($z$) state at $J'/J = 0.2$ and $\Delta = 1$ and 
(b) the AFQ(${\varv}$) state at $J'/J = 0.7$ and $\Delta = 0.5$.
The insets of the (a) and (b) are the temperature dependences of the order parameters.
Other parameters are fixed at $g^{\Gamma_6} = -0.4$, $g^{\Gamma_7} = 0.5$, and $g' = 0.8$.
 \label{fig:finite_temperature}
}
\end{figure}

Figure~\ref{fig:finite_temperature}(a) shows 
$\alpha_{xx}^{\rm (E)}$ as a function of temperature $T$ in the AFM($z$) phase with the magnetic quadrupole $M_u$ 
(finite $\alpha_{xx}^{\rm (E)}=\alpha_{yy}^{\rm (E)}$ and $\alpha_{zz}^{\rm (E)}$) at $J'/J=0.2$, $\Delta=1$, $g^{\Gamma_6}=-0.4$, $g^{\Gamma_7}=0.5$, and $g' = 0.8$~\cite{comment_ME_intra}.
$\alpha_{xx}^{\rm (E)}$ becomes nonzero below $T_{\rm N}\simeq 0.77$ 
and decreases for $0.69 \lesssim T \lesssim 0.75$ after showing the peak structure at $T\simeq 0.75$. 
While further decreasing $T$, $\alpha_{xx}^{\rm (E)}$ grows and becomes the largest at the lowest $T$.
The complicated temperature dependence of $\alpha_{xx}^{\rm (E)}$ is due to the orbital degree of freedom.  
Its qualitative behavior is characterized by each component $\alpha_{xx}^{({\rm E}, \Gamma_6)}$, $\alpha_{xx}^{({\rm E}, \Gamma_7)}$, and $\alpha_{xx}^{\prime {\rm (E)}}$, as also plotted in Fig.~\ref{fig:finite_temperature}(a).
$\alpha_{xx}^{\rm (E,\Gamma_7)}$ increases with onset of $M^{\rm AF}_{z}$ in the inset of Fig.~\ref{fig:finite_temperature}(a), since $M_z^{\rm AF}$ mainly consists of the magnetic moment in the $\Gamma_7$ orbital.
On the other hand, 
as a further increase of $M^{\rm AF}_{z}$ leads to the large energy gap between the up- and down-spin bands of the $\Gamma_7$ orbital, $\alpha_{xx}^{\rm (E,\Gamma_7)}$ decreases and the interorbital contribution, $\alpha_{xx}^{\prime {\rm (E)}}$, becomes dominant for $T \lesssim 0.69$.  
It means that the interorbital $M_x^\prime$ activated in $\Gamma_6$-$\Gamma_7$ space
is significant for the large magneto-electric response in this multi-orbital system. 
The typical magnitude of the magneto-electric tensor is estimated as $\sim10^{-1}|t_{\parallel}^{\Gamma_7}|^{-1}$ ps m$^{-1}$ in the unit of $|t_{\parallel}^{\Gamma_7}|$~eV.

We show nonzero $\alpha_{yx}$ in the AFQ($\varv$) phase in Fig.~\ref{fig:finite_temperature}(b)
 in addition to the order parameter $Q_{\varv}^{{\rm AF}}$ in the inset of Fig.~\ref{fig:finite_temperature}(b) at $J'/J = 0.7$, $\Delta = 0.5$, $g^{\Gamma_6}=-0.4$, $g^{\Gamma_7}=0.5$, and $g'=0.8$.
We find the finite-temperature phase transition between AFQ($\varv$) state and the antiferrooctupole (AFO) state with $M_{xyz}^{{\rm AF}}$ at $T_{0} \sim 0.75$. 
From the symmetry in Table~\ref{table:cross_correlation}, in the former AFQ state, odd-parity $G_{xy}$ induces $\alpha_{yx}^{\rm (J)}=\alpha_{xy}^{\rm (J)}$, while
$M_{xy}$ in the AFO state shows $\alpha_{yx}^{\rm (E)}=\alpha_{xy}^{\rm (E)}$. 

In the AFQ$(\varv)$ state in Fig.~\ref{fig:finite_temperature}(b), the amplitude of $\alpha_{yx}^{\rm (J)}$ increases from the lowest $T$ and it shows the peak at $T \sim 0.25$, where $Q_{\varv}^{\rm AF}$ reaches almost full saturation.   
While further increasing $T$, $|\alpha_{yx}^{\rm (J)}|$ gradually decreases and jumps at the phase boundary with the AFO state.
The temperature dependence of $\alpha_{yx}^{\rm (J)}$ reflects the electronic state around the Fermi surface, since the intraband contribution is dominant.
For $\alpha_{yx}^{\rm (J)}$ in the AFQ($\varv$) state, the interorbital component $\alpha_{yx}^{\prime{\rm (J)}}$ becomes dominant, while $\alpha_{yx}^{\rm (J,\Gamma_6)}$ and $\alpha_{yx}^{\rm (J,\Gamma_7)}$ almost cancel with each other, which also shows that the interorbital component $\alpha_{yx}^{\prime{\rm (J)}}$ is significant in  this multi-orbital system.
The typical magnitude of $\alpha_{yx}^{\rm (J)}$ in the AFQ($\varv$) state is estimated as $\sim10^{-1}|t_{\parallel}^{\Gamma_7}|^{-1}\delta^{-1}$ ps m$^{-1}$ for $|t_\parallel^{\Gamma_7}|$~eV and the broadening factor $\delta$~s$^{-1}$. 
On the other hand, the electric conductivity is obtained as $10^{-3}\delta^{-1}$~$\mu \Omega^{-1}$cm$^{-1}$, which implies $\delta \sim 10^{-2}$-$10^{-1}$ from the comparison with the experimental data~\cite{tanida2019successive}.
Therefore, the large magneto-electric response might be expected in CeCoSi.

As nonzero components of $\alpha_{\mu\nu}$ are the same for the AFQ($\varv$) and AFO states, 
it is difficult to distinguish them in terms of the cross-correlation measurements. 
Meanwhile, the difference is in the electronic band structure. In the AFQ($\varv$) state, the electronic band structure shows the spin splitting as the functional form of $k_x \sigma_y + k_y \sigma_x$, while there is not antisymmetric spin splitting in the AFO state~\cite{hayami2018classification}.
Thus, the spin- and angle-resolved photoemission spectroscopy is also a powerful tool to identify order parameters.

To summarize, we investigate the possibility of odd-parity multipoles induced by the AFM and AFQ orderings in CeCoSi. 
We analyze the two-orbital model including the CF splitting between the ground-state $\Gamma_7$ level and the first-excited $\Gamma_6$ level and the staggered ASOI by the mean-field calculations.
We show that the AFM state with $3z^2-r^2$ type of the magnetic quadrupole and the AFQ state with $xy$ type of the electric toroidal quadrupole are dominantly stabilized by changing the CF splitting, multipole-multipole interactions, and the staggered ASOI. 
Furthermore, we also discuss the behavior of the magneto-electric tensor driven by odd-parity multipoles.
Although model parameters used in the present study need to be more sophisticated by first-principle calculations, our analysis will serve as a starting point for identifying order parameters in CeCoSi.

We thank H. Tanida and K. Mitsumoto for the fruitful discussions on experimental information in CeCoSi. This research was supported by JSPJ KAKENHI Grant Numbers JP18H04296 (J-Physics) and 19K03752.

\bibliographystyle{jpsj}
\bibliography{17611}

\begin{thebibliography}{10}
\bibitem{comment_interorbital}
{The intraorbital and interorbital ASOIs do not stabilize the states with
  ($M'_x$, $M'_y$) and ($Q_{yz}$, $Q_{zx}$) in the present model, which is
  consistent with the results obtained from the effective Hamiltonian in the
  strongly correlated limit (not shown).}
\bibitem{comment_ME_intra}
{We here omit the result of $\alpha_{zz}^{\rm (E)}$, as the magnitude of
  $\alpha_{zz}^{\rm (E)}$ is smaller than that of $\alpha_{xx}^{\rm (E)}$ by
  the order of $10^{-2}$ because of the low conductivity in the $z$ direction.}
\end{thebibliography}

\end{document}